# Förster energy transfer boosts indirect anisotropic interlayer excitons in 2L-$MoSe_2$/perovskite heterostructures

Yingying Chen (陈瑛瑛)[1], Zihao Jiao (焦子豪)[2], Haizhen Wang (王海珍)[2] and Dehui Li (李德慧)[2,3*]

*1 School of Science, Jimei University, Yinjiang Road 185, Xiamen 361021, China*

*2 School of Optical and Electronic Information, Huazhong University of Science and Technology, Luoyu Road 1037, Wuhan 430074, China*

*3 Wuhan National Laboratory for Optoelectronics, Huazhong University of Science and Technology, Luoyu Road 1037, Wuhan 430074, China*

*Correspondence to: Email: dehuili@hust.edu.cn.

**Abstract**

Interlayer excitons (IXs) in two-dimensional (2D) van der Waals heterostructures have attracted considerable attention due to their unique optical and electronic properties. Owing to the spatially indirect nature, the radiative emission efficiency highly sensitive to interlayer twist angles. Further considering that their uniformly oriented out-of-plane dipole moments limit directional emission, strategies to simultaneously improve emission efficiency and induce optical anisotropy warrant in-depth investigation. In this work, we report significant photoluminescence (PL) enhancement and optical anisotropy of IXs in 2L-$MoSe_2$/perovskite heterostructures mediated by energy transfer from $ReS_2$. We attribute this enhancement to Förster resonance energy transfer (FRET), which increases the 2L-$MoSe_2$ emission by approximately eight-fold at room temperature, and nearly doubles the emission intensity of momentum-indirect IXs in 2L-$MoSe_2$/perovskite heterostructures at 78 K. Importantly, the optical anisotropy of $ReS_2$ can be effectively imprinted onto 2L-$MoSe_2$ and associated indirect IXs during the energy transfer process, yielding a linear dichroism of approximately 1.1 for both intralayer excitons and IXs with identical polarization directions. These findings expand the scope of IX study beyond direct bandgap materials with strong intrinsic emission to include systems with indirect bandgaps, offering new avenues for realizing high-performance polarization-sensitive optoelectronic devices.

## Introduction

Energy transfer and charge transfer constitute the fundamental basis of van der Waals heterostructures, giving rise to fascinating phenomena and novel physical states[1-3]. Due to the reduced dielectric screening in layered materials, exciton states formed by charge transfer or modulated by energy transfer dominate the emission characteristics of these materials and their heterostructures[4]. These exciton states exhibit distinct luminescence properties across different systems[5-7]. For instance, monolayer transition metal dichalcogenides (TMDs) with direct bandgaps exhibit strong exciton emission with large binding energies that can sustain to room temperature[8-9]. In contrast, the emission efficiencies of few-layer TMDs possessing indirect bandgaps, as well as their heterostructures featuring indirect interlayer transitions, are generally characterized by low radiative emission efficiency[10-12]. Therefore, improving their emission performance and efficiency are of significant research importance. In this situation, many approaches have been be implemented to boost their performance, such as chemical treatment[13], dielectric and strain engineering[14-15], plasmonic enhancement[16], cavity integration[17], electrostatic doping[18], and introducing strong light-matter interactions to form polaritons[19], moiré excitons[20], or single photon emitters[21]. Among these, energy transfer, referring to donor-acceptor exchange electronic excitation, appears highly competitive owing to its lossless and low-cost nature[3]. This phenomenon has been extensively discussed in quantum dots system, and subsequently extends to van der Waals (vdW) heterostructures[22-23]. Generally, there are two types of energy transfer according to their different working mechanism, including Förster type which relies on dipole-dipole coupling over long distances (~10 nm), and Dexter type which involves electron exchange over short distances (~1 nm). These energy transfer processes have been demonstrated to efficiently improve their emission performance, which can be further modulated by external electric field[24], magnetic field[25], electrostatic gating[26], strain[15, 27] or special intermediate states such as dark[28] or entangled states[29]. These findings broaden the scope of optoelectronic applications in light-emitting diodes[30], solar cells[31], photodetectors[32] and phototransistors[33].

In particular, interlayer excitons (IXs) in vdW heterostructures, characterized by spatially indirect natures exhibiting long lifetimes and repulsive dipolar interaction, have attract intensive attention[34]. Recently, several studies have demonstrated the enhanced IX emission in TMD-based heterostructures[35-36]. Most investigations have concentrated on momentum-direct IXs or those with small twist angles in TMD heterobilayers that already possesses strong radiative emission, while momentum-indirect IXs in few-layer TMD-based heterostructures remain rarely discussed. Moreover, studies have shown that heterostructures incorporating other 2D materials, e.g. InSe or $ReS_2$, facilitates the transfer of optical anisotropy to IXs in addition to enhance their emission intensity by tailoring the energy transfer process[37-40]. These heterostructures exhibit strong anisotropic IX emission variations, therefore, by combining and utilizing these functional components, IXs in these systems are expected to yield novel phenomena that may attract renewed attention. Besides, considering materials with inherently weak light absorption, e.g. $MoSe_2$, it's of great importance to boost their emission to achieve effective optoelectronic devices.

Here, we report enhanced IX emission with induced optical anisotropy in $ReS_2$/hBN/2L-$MoSe_2$/perovskite heterostructures. By utilizing few-layer $ReS_2$, the exciton emission can be strongly boosted with an enhancement factor of approximately eight-fold for 2L-$MoSe_2$ even at room temperature. Such enhancement further extends to indirect IXs in 2L-$MoSe_2$/(iso-BA)$_2$PbI$_4$ heterostructure, showing a maximum enhancement factor of approximately two-fold at liquid nitrogen temperature. This enhancement capability exhibits varying power and temperature dependence between intralayer excitons in 2L-$MoSe_2$ and IXs in 2L-$MoSe_2$/(iso-BA)$_2$PbI$_4$, contingent upon the energy transfer efficiency. More importantly, the optical anisotropy of $ReS_2$ has been demonstrated to transfer successfully to 2L-$MoSe_2$ and resultant IXs with identical polarization direction and a linear dichroism of approximately 1.1. These observations facilitate the generation of momentum-indirect IXs with both strong emission and anisotropy, circumventing the strict requirement for specific rotation angles in TMD heterobilayers to form near-direct interlayer transitions, thereby expanding their potential applications for enhancing the indirect transitions in next-generation optoelectronic devices.

**Results and Discussion**

Figure 1a illustrates the schematic diagram of the $ReS_2$/hBN/2L-$MoSe_2$ heterostructure, in which the few-layer $ReS_2$ and 2L-$MoSe_2$ are separated by a few-layer hBN flake. The heterostructure was fabricated by mechanical exfoliation from bulky crystals followed by the dry-transfer method onto a $SiO_2$ (300 nm)/Si substrate (Fig. 1b). According to previous literature, the band alignment between few-layer $ReS_2$ and 2L-$MoSe_2$ forms an indirect type-II configuration, favoring both energy and charge transfer processes[41-43]. To eliminate the effect of charge transfer process which would quench their PL intensity, a thin layer of hBN (~10 nm) was inserted to inhibit charge transfer while permitting the occurrence of energy transfer. Figure 1c displays the PL spectra of the $ReS_2$/hBN/2L-$MoSe_2$ heterostructure at room temperature. While $ReS_2$ shows an exciton peak at 790 nm in the isolated region but becomes invisible in the heterostructure region, 2L-$MoSe_2$ exhibits an obvious enhancement around 810 nm in the heterostructure region. It should be noted that 2L-$MoSe_2$ exciton also shows a stronger emission in the hBN/2L-$MoSe_2$ region compared to bare 2L-$MoSe_2$, mainly due to the suppression of surface disorder and defects induced between the 2L-$MoSe_2$ and $SiO_2$/Si substrate[44]. Therefore, to avoid ambiguity regarding substrate influence, we focus our comparison on the PL intensity between hBN/2L-$MoSe_2$ and $ReS_2$/hBN/2L-$MoSe_2$. Notably, the indirect intralayer exciton emission shows the strongest PL intensity in the $ReS_2$/hBN/2L-$MoSe_2$ heterostructure. Combing the hBN insertion and the observed quenched $ReS_2$ emission, this enhanced emission in this heterostructure can be attributed to the non-radiative energy transfer from few-layer $ReS_2$ to 2L-$MoSe_2$, facilitated by the significant spectral overlap between the $ReS_2$ absorption and 2L-$MoSe_2$ emission[45-46]. Such energy transfer should be ascribed to the Förster type according to our previous study in the $ReS_2$/hBN/1L-$MoSe_2$ device[47], in which the ~10 nm thickness of hBN spacer allows optimal and efficient dipole-dipole coupling[48] while effectively eliminating the possibility of Dexter type energy transfer or charge tunneling. Notably, although 2L-$MoSe_2$ possesses a momentum-indirect bandgap, its direct and indirect bandgap values are almost degenerate[49]. At finite temperatures, due to the layer decoupling caused by interlayer thermal expansion[45], the direct K−K transition dominates its exciton emission in the PL spectra. Consequently, the Förster resonance energy transfer (FRET) from $ReS_2$ preferentially populates these direct exciton states, thereby boosting the observed exciton emission intensity. Moreover, this PL enhancement differs from local interfacial charge transfer or heterostructure inhomogeneity since the uniform enhanced emission across the heterostructure region has been demonstrated in our previous work[47].

To quantify the PL enhancement, we define the PL enhancement factor as $R = I_{HS}/I_M$, where $I_{HS}$ is the PL intensity of $ReS_2$/hBN/2L-$MoSe_2$ heterostructure and $I_M$ represents the constituent reference hBN/2L-$MoSe_2$ emission intensity. Here, we compare $ReS_2$/hBN/2L-$MoSe_2$ with hBN/2L-$MoSe_2$ but not bare 2L-$MoSe_2$ to eliminate the influence of dielectric environment variation. All samples share the same hBN dielectric interface and the same hBN/$SiO_2$ dielectric condition underneath. Interestingly, we observe that this enhancement factor decreases with increasing excitation power. As shown in Figure 1d, 2L-$MoSe_2$ emission shows pronounced enhanced PL intensity under the excitation power of 82.8 μW and moderate enhancement with the power of 8020 μW. We further perform the power-dependent PL studies to elucidate the in-depth energy transfer mechanism. While emission intensities of both hBN/2L-$MoSe_2$ and $ReS_2$/hBN/2L-$MoSe_2$ increasing with increasing incident power, the rate of increase for $ReS_2$/hBN/2L-$MoSe_2$ is notably smaller (Figure 1e). Further fitting PL intensity with incident power using the power law $I = P^{\alpha}$, where $I$, $P$ and $\alpha$ donates the PL intensity, laser power and power index, respectively. We determined that the $\alpha$ value of 2L-$MoSe_2$ emission in $ReS_2$/hBN/2L-$MoSe_2$ and hBN/2L-$MoSe_2$ regions are 0.6 and 0.9, respectively. The detailed power-dependent PL comparisons are provided in Fig. S1. Although lasing heating effects may account for some reason for this behavior at high power range above 1 mW, the energy transfer plays an important role at elevated powers. Figure 1f summarizes the enhancement factor trend across this excitation power range, which shows a maximum R≈8.2 at 82.8 μW and minimum R≈2.2 at 8020 μW. This decreased enhancement may arise from the intense energy transfer that inducing large exciton density, leading to exciton-exciton annihilation at sufficiently high excitation powers thus reducing $\alpha$ value. These observations further validate the contribution of FRET to this PL enhancement and rule out the effect of dielectric environment variations as it should be power-independent. We also note that such enhancement factor is smaller than that in our previously reported $ReS_2$/hBN/1L-$MoSe_2$ heterostructure which exhibits a ~58.7-fold enhancement[47]. The smaller spectral overlap between the absorption of 2L-$MoSe_2$ and the emission of $ReS_2$, as well as the indirect nature of 2L-$MoSe_2$, likely contribute to this moderate energy transfer efficiency and PL enhancement. Nevertheless, FRET remains an effective and non-destructive method for addressing weak emission especially in indirect system, which can improve radiative emission efficiency and yield a large relative enhancement factor.

Interestingly, further stacking a thin flake of 2D perovskite layer can induce intriguing phenomena (Fig. 2a). As we observed in our previous work, TMD/perovskite heterostructures can form robust IXs irrespective of stacking angles and lattice mismatch[50-51]. In this case, the IX emission in the TMD/perovskite heterostructure is expected to be efficiently boosted through the energy transfer from $ReS_2$. To validate this assumption, we fabricated a $ReS_2$/hBN/2L-$MoSe_2$/(iso-BA)$_2$$PbI_4$ heterostructure using the same assembly methods (iso-BA represents iso-butylamine with a chemical formula of $C_4H_9NH_3$) (Fig. 2b). As shown in Fig. 2c, 2L-$MoSe_2$ and (iso-BA)$_2$$PbI_4$ can form type-II heterostructures with a momentum-indirect nature, enabling the formation of indirect IXs, consistent with our previous work[11, 50]. Since the inserted hBN layer allows energy transfer from $ReS_2$ to 2L-$MoSe_2$/(iso-BA)$_2$$PbI_4$ heterostructure (hereafter referred as HS) and blocks charge transfer from $ReS_2$ to HS, such energy transfer is anticipated to manipulate the optical performance of IX emission.

We subsequently carried out PL measurements at 78 K, as shown in Fig. 2d, while hBN/HS shows a peak around 775 nm corresponding to the intralayer exciton of 2L-$MoSe_2$, $ReS_2$/hBN/HS exhibit an additional peak near 795 nm and a strong peak around

777 nm that can be ascribed to the superposition of emission from few-layer $ReS_2$ and 2L-$MoSe_2$[45, 52]. Similar phenomena can be observed in the $ReS_2$/hBN/2L-$MoSe_2$ heterostructure, which shows enhanced overall emission intensity due to the efficient energy transfer (Fig. S2). Importantly, at the low energy side, a pronounced broad peak appears with a central peak at 925 nm, which can be ascribed to IXs in hBN/HS according to their band alignment[41, 43]. Notably, this IX emission exhibit almost two-fold PL enhancement in the presence of $ReS_2$-integrated heterostructure. We attribute this enhanced emission to energy transfer from $ReS_2$ to the heterostructure, enabling momentum-indirect IXs in $ReS_2$/hBN/HS with increased emission intensity. Specifically, considering the in-plane exciton nature of $ReS_2$ and out-of-plane nature of IXs, the energy and charge transfer process can be described as follows: (1) initial exciton generation in 2L-$MoSe_2$ via FERT from $ReS_2$ to 2L-$MoSe_2$ to form intralayer hot excitons, predominantly populating the direct states with a fraction evolving into the indirect states; (2) rapid intralayer thermalization and intervalley scattering in 2L-$MoSe_2$; (3) charge transfer occurs at the type-II 2L-$MoSe_2$/(iso-BA)$_2$Pb$I_2$ interface to form momentum-indirect IXs. Since the hBN layer hinders the direct charge transfer from $ReS_2$ to HS, FRET provides more energy to 2L-$MoSe_2$ to increase the exciton population, thus enhancing the subsequent IX emission.

To support this hypothesis, we performed power-dependent PL studies under 633 nm laser excitation at 78 K. As shown in Fig. 2e, the power dependence of IXs in $ReS_2$/hBN/HS and hBN/HS exhibit divergent tendencies with a turning point around the power of 200 μW. While IXs in hBN/HS show a uniform slope around 0.5, IXs in $ReS_2$/hBN/HS exhibit a larger slope of ~0.6 below the power of 200 μW and a smaller slope of ~0.4 above the power of 200 μW. Furthermore, the peak wavelength of IXs also shows a large shift range of 43 nm in $ReS_2$/hBN/HS, significantly larger than the 18 nm shift range observed for IXs in hBN/HS (Fig. S3). We further calculated the enhancement factor by comparing the PL intensity between $ReS_2$/hBN/HS and hBN/HS in Figure 2f. As expected, the enhancement factor initially rises with increasing incident power, reaching a maximum enhancement factor of ~1.8 under the power of 200 μW, followed by a decrease at elevated powers. This power dependence contrasts

significantly with 2L-$MoSe_2$ exciton emission shown in Fig. 1f, but exactly explains the different slope values in Fig. 1e. Since $ReS_2$/hBN/2L-$MoSe_2$ shows largely enhanced emission intensity under low power conditions, the IX density in $ReS_2$/hBN/HS increase more rapidly with a large slope value. Conversely, the saturation of PL enhancement in $ReS_2$/hBN/2L-$MoSe_2$ at high power regimes results in a reduced slope value for IXs in $ReS_2$/hBN/HS. We should note that this enhancement factor is smaller than that of $ReS_2$/hBN/2L-$MoSe_2$, but still comparable to IXs in the previously reported $ReS_2$/hBN/1L-$MoSe_2$/(iso-BA)$_2$$PbI_4$ device[47] and other $ReS_2$/$MoS_2$ heterostructure[40]. Besides, this observation cannot be simply attributed to the heterostructure inhomogeneity as the interface defect states should not exhibit this unique power-dependent tend.

To further explore the energy transfer efficiency and its influence on PL enhancement, we carried out the temperature-dependent PL studies of both $ReS_2$/hBN/2L-$MoSe_2$ and $ReS_2$/hBN/HS. As show in Fig. S4, the exciton peak in $ReS_2$/hBN/2L-$MoSe_2$ gradually redshifts with increasing temperature, consistent with bandgap narrowing in conventional semiconductors. Interestingly, due to relatively close exciton emission in few-layer $ReS_2$ and 2L-$MoSe_2$, these two exciton peaks become distinguishable only at low temperature (Fig. S2). As temperature increases, due to the energy transfer from $ReS_2$ to 2L-$MoSe_2$ depending on the original PL intensity of $ReS_2$ and the absorption-emission spectral overlap between $ReS_2$ and 2L-$MoSe_2$, the exciton emission exhibit contrasting enhancement behavior. Specifically, the PL intensities of both hBN/2L-$MoSe_2$ and $ReS_2$/hBN/2L-$MoSe_2$ initially decreases with increasing temperature, reaching a minimum at 200 K, before increasing up to room temperature (Fig. 3a), coinciding with the sudden reduction in spectral overlap ratio around 200 K, similar to that in $ReS_2$/hBN/1L-$MoSe_2$ reported before. This observation also suggests that although enhanced phonon-assisted indirect transitions increase non-radiative recombination with increasing temperature in 2L-$MoSe_2$, efficient energy transfer increase the radiative recombination rate, enabling increased PL intensity of 2L-$MoSe_2$ at elevated temperatures.

By comparing the PL intensity of $ReS_2$/hBN/2L-$MoSe_2$ with that of hBN/2L-$MoSe_2$, we can obtain the PL enhancement factor as a function of temperature, as shown in Fig. 3b. This enhancement factor gradually increases with increasing temperature, but shows a sudden drop at 200 K, followed by a continuous increase to a maximum value at room temperature. These observations indicate that the PL enhancement in the $ReS_2$/hBN/2L-$MoSe_2$ heterostructure indeed originates from energy transfer from $ReS_2$ to 2L-$MoSe_2$, which is related to the energy transfer efficiency relaying on the spectral overlap between $ReS_2$ and 2L-$MoSe_2$ and total energy of the donor and acceptor system. We should note that this PL enhancement in $ReS_2$/hBN/2L-$MoSe_2$ at low temperature is comparatively moderate relative to their monolayer $MoSe_2$-based counterparts. This phenomenon may be attributed to three reasons: 1) the small bandgap of 2L-$MoSe_2$ reduces the overlap between its absorption spectrum and the emission spectrum of $ReS_2$, resulting a lower the energy transfer efficiency; 2) the indirect bandgap nature at low temperatures limits the intrinsic emission efficiency of 2L-$MoSe_2$ itself, which further impacts the energy transfer efficiency in the $ReS_2$/hBN/2L-$MoSe_2$ heterostructure; 3) both 2L-$MoSe_2$ and few-layer $ReS_2$ require phonon-assisted transition upon excitation, thereby complicating the energy transfer process. Nevertheless, despite these limitations, this energy transfer mechanism effectively improves the emission efficiency of the indirect semiconductors.

In contrast, IXs in $ReS_2$/hBN/HS exhibit completely opposite temperature dependence. As shown in Fig 3c, the IX emission intensity decreases monotonically with increasing temperature for both $ReS_2$/hBN/HS and hBN/HS. This observation can be ascribed to the gradually weakened interlayer coupling, leading to reduced electron-hole wave function overlap, thereby decreasing radiative recombination rate and resulting in low emission intensity at elevated temperatures, which becomes distinguishable above 220 K. We also calculated the enhancement factor as function of temperature. As shown in Fig. 3d, this enhancement factor maintains a relatively constant value at approximately1.1 across varying temperature. We should note that these R values were acquired under the laser power of 2 mW to ensure visible IX emission due to the significant reduction in indirect IX emission at elevated temperatures. As indicated in Fig. 2f, the enhancement factor decreases as the incident power exceeds 200 μW; therefore, a smaller enhancement factor is observed in Fig. 3d. Nevertheless, this result does not affect our observation of the enhancement factor on temperature dependence. It is worth noting that this constant PL enhancement contrasts significantly with the increasing trend observed in $ReS_2$/hBN/2L-$MoSe_2$, which can be attributed to the interplay between varying energy transfer efficiency and band alignment with increasing temperature. As show in Fig. S5, the IX emission initially blueshifts as temperature increases to 140 K, and then redshifts. This behavior contradicts the expectations of a simple reduction in interlayer bandgap as previously reported[50], which should be explained by the shifting conduction band minimum of 2L-$MoSe_2$ from Γ−K to K point with increasing temperature[43]. Below 140 K, the shift drives the blueshift; above 140 K, as the conduction band minimum of 2L-$MoSe_2$ settles at the K point, the IX peak energy follows the standard bandgap reduction and thereby redshifts with further increasing temperature. The combined effect of these complex factors renders the energy transfer efficiency in $ReS_2$/hBN/HS relatively insensitive to temperature variations.

Furthermore, considering the anisotropic nature of $ReS_2$ which shows significant differences in light absorption and carrier concentration along different directions, we speculated that the optical anisotropy should be transmitted to IXs accompanied by energy transfer in $ReS_2$/hBN/HS. To validate this, we performed the polarization-resolved PL measurement by rotating a λ/2 waveplate in the excitation path to vary the polarization direction of the excitation light. The emission intensity was then collected and plotted as an anisotropic PL polar plot, as shown in Fig.4. Figure 4a displays a significant anisotropy of $ReS_2$, showing a linear dichroism (defined as $I_{max}/I_{min}$, where $I_{max}$ and $I_{min}$ represent the dichroic fitted maximum and minimum intensity, respectively) of approximately 2.1. In contrast, due to the intrinsic in-plane isotropic nature of 2L-$MoSe_2$, hBN/2L-$MoSe_2$ shows uniform emission intensity that is independent of incident polarization (Fig.4b). Remarkably, both $ReS_2$/hBN/2L-$MoSe_2$ and $ReS_2$/hBN/HS exhibit significant optical anisotropy with an identical linear dichroism of about 1.1. This demonstrates the successful anisotropy transfer to both intralayer excitons and IXs, as the intrinsic IXs should be isotropic given the isotropic nature of 2L-$MoSe_2$ and (iso-BA)$_2$$PbI_4$. In this situation, we believe that the anisotropy transfer originates primarily from anisotropic absorption in $ReS_2$ to modulate this FRET process and associated exciton density. The mechanism can be described as follows: (1) anisotropic light absorption in $ReS_2$ which shows strongest along Re-Re chain directions to introduce an initial polarization-dependent exciton population; (2) a large amount energy transfer along the Re-Re chain direction from $ReS_2$ to 2L-$MoSe_2$; (3) exciton population modulation in 2L-$MoSe_2$ via FRET from $ReS_2$ despite its isotropic nature; (4) anisotropic IX emission from the preserved population imbalance, which achieves a higher IX density along the Re chains direction, thus resulting in net anisotropy transfer. The energy and charge transfer process can also be seen in Fig. S6.

It is also noteworthy that the polarization direction is identical in $ReS_2$/hBN/2L-$MoSe_2$ and $ReS_2$/hBN/HS, but differs from that of individual $ReS_2$, which can also be seen in another $ReS_2$/hBN/2L-$MoSe_2$ sample (Fig. S7). Nevertheless, $ReS_2$/hBN/2L-$MoSe_2$ and $ReS_2$/hBN/HS share identical polarization directions, which may result from a large carrier population along this specific direction in 2L-$MoSe_2$, and thereby promoting intralayer exciton dissociation upon excitation and subsequently facilitating IX formation in 2L-$MoSe_2$/(iso-BA)$_2$$PbI_4$. Although this linear dichroism decreases from ~2.1 in $ReS_2$ to ~1.1 in IXs that may be ascribed to potential depolarization through energy transfer and indirect IX formation process (such as intervalley scattering, phonon interactions, and etc)[47], this value is robust and comparable to IXs that reported in other $ReS_2$-based structures[40, 53]. These observations suggest that energy transfer serves as an effective strategy to imprint optical anisotropy, paving the way to polarization-sensitive optical devices.

## Conclusion

In summary, we have demonstrated the efficient energy transfer from $ReS_2$ to 2L-$MoSe_2$ and 2L-$MoSe_2$/(iso-BA)$_2$$PbI_4$ heterostructures, achieving anisotropic IXs with enhanced emission intensity in $ReS_2$/hBN/2L-$MoSe_2$/(iso-BA)$_2$$PbI_4$ heterostructures. We first established the presence of FRET in $ReS_2$/hBN/2L-$MoSe_2$, in which the PL intensity can be enhanced eight-fold at room temperature. This efficient energy transfer subsequently amplifies momentum-indirect IX emission in 2L-$MoSe_2$/(iso-BA)$_2PbI_4$ heterostructures by nearly two-fold, enabling a distinct linear dichroism around 1.1. These observations indicate that optical anisotropy can be effectively transmitted through energy transfer in the heterostructure assembly of few-layer TMDs and 2D perovskites, enabling the realization of anisotropic exciton emission incorporated with substantially enhanced intensity. Our findings also evidence that FRET provide a solid platform for enhancing momentum-indirect transitions, offering a robust strategy for realizing highly efficient anisotropic sources in next-generation optoelectronics.

**Methods**

**Sample preparations:** $ReS_2$, $MoSe_2$ and hBN are bulk crystals were purchased from HQ Graphene while (iso-BA)$_2PbI_4$ bulk crystals were synthesized by solution methods according to our previous work. All constituent flakes were mechanically exfoliated from their bulky crystals on polydimethylsiloxane **(**PDMS) stamp and then stacked layer by layer on $SiO_2$/Si substrate by dry transfer methods.

**Optical measurements**: The optical images were captured on an Olympus microscope (BX53M). Samples were mounted in a liquid nitrogen bath cryostat (Cryo Industries of America Inc.) and were measured on a home-built Raman spectrometer (Horiba iHR-550) with a 600 g/mm grating. The PL spectra were acquired under the excitation of a 633-nm He-Ne laser with a 50× objective lens (NA = 0.65) under varying temperatures, which can be continuously tuned from 78 K to 300 K by a temperature controller (Lake Shore Cryotronics, Model 336). For polarization-resolved PL measurement, a set of polarizer (Thorlabs WP25M-VIS) and half-wave plate (Thorlabs AHWP05M-600) were used to control the polarization direction of incident laser, and a polarizer (Thorlabs WP25M-VIS) was placed in the collection optical path to eliminate the polarization of response grating and CCD.

**Supporting Information**

Supporting Information is available online or from the author.

**Acknowledgments**

The authors acknowledge the support from the National Key Research and Development Program of China (2024YFA1208500), the Hubei Provincial Natural Science Foundation (2025AFA039) and Open Project Program of Hubei Optical Fundamental Research Center (HBO2026C015), the Natural Science Foundation of Fujian province (2025J08202) and Xiamen City (3502Z202471047).

**ORCID iDs**

Yingying Chen https://orcid.org/0000-0003-4519-5836

Dehui Li https://orcid.org/0000-0002-5945-220X

**Conflict of Interest**

The authors declare no conflict of interests.

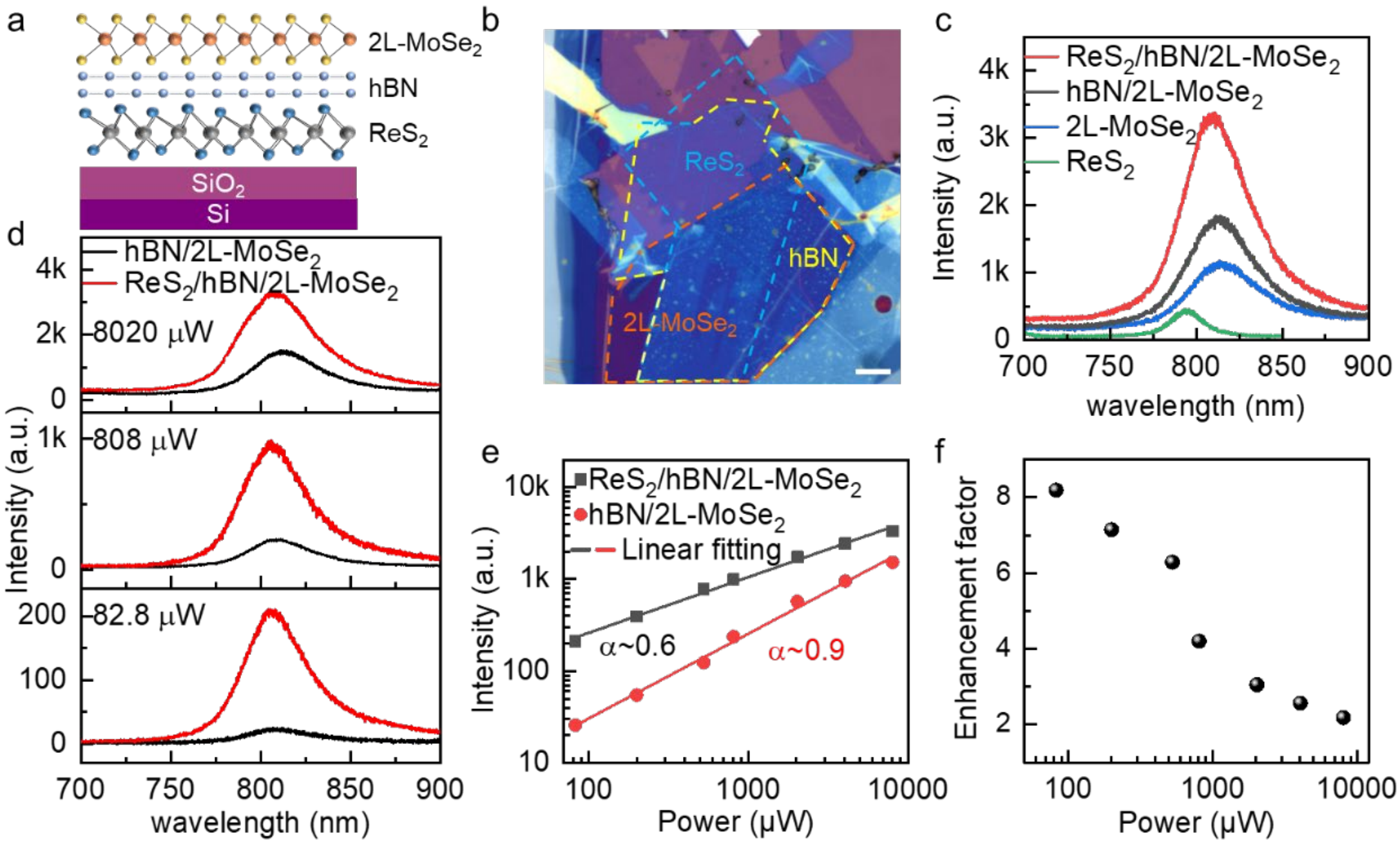

**Figure 1. Energy transfer from $ReS_2$ to 2L-$MoSe_2$ to boost emission.** Schematic illustration (a) and optical image (b) of the $ReS_2$/hBN/2L-$MoSe_2$ heterostructure on the $SiO_2$/Si substrate. The dashed orange, yellow and blue lines outline the edge of 2L-$MoSe_2$, few-layer hBN and multi-layer $ReS_2$, respectively from top to bottom. Scale bar, 10 μm. (c) PL spectra of the $ReS_2$/hBN/2L-$MoSe_2$ heterostructure and constituent components at room temperature under the excitation of 633 nm laser with a power of 800 μW. (d) PL comparison of hBN/2L-$MoSe_2$ and $ReS_2$/hBN/2L-$MoSe_2$ heterostructures under different excitation power of 633 nm laser. (e) Power-dependent emission intensity of hBN/2L-$MoSe_2$ and $ReS_2$/hBN/2L-$MoSe_2$ heterostructures and linear fittings. (f) Power dependence of enhancement factor in the $ReS_2$/hBN/2L-$MoSe_2$ heterostructure.

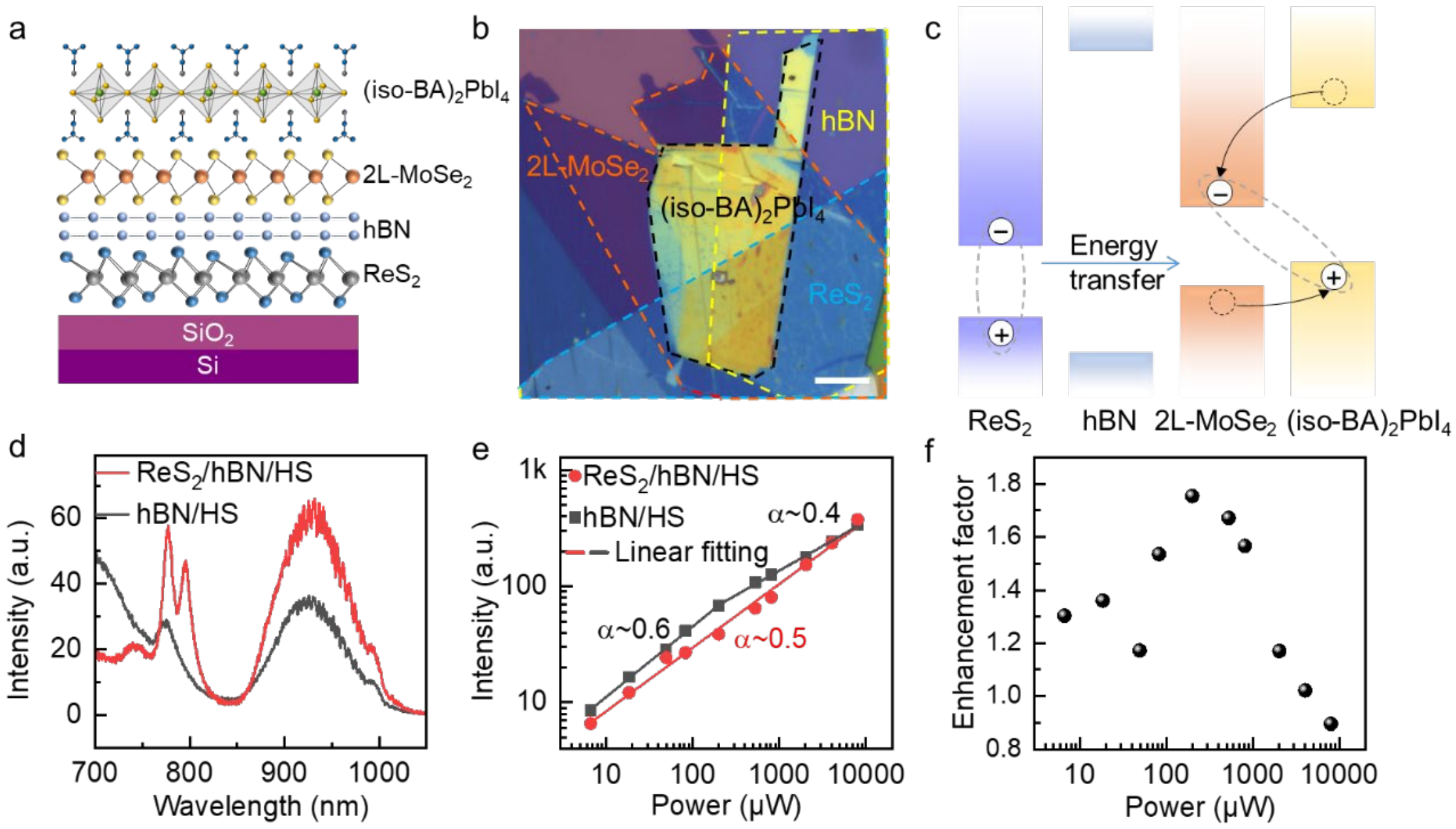

**Figure 2. Enhanced IX emission in $ReS_2$/hBN/2L-$MoSe_2$/(iso-BA)$_2PbI_4$ heterostructure via energy transfer.** Schematic illustration (a) and optical image (b) of the $ReS_2$/hBN/2L-$MoSe_2$/(iso-BA)$_2PbI_4$ heterostructure. The dashed black, orange, yellow and blue lines outline the edge of (iso-BA)$_2PbI_4$ flake, 2L-$MoSe_2$, few-layer hBN and multi-layer $ReS_2$ from top to bottom. Scale bar, 10 μm. (c) Schematic band diagram of the dynamic processes, including the charge transfer process in 2L-$MoSe_2$/(iso-BA)$_2PbI_4$ to form IXs and energy transfer from $ReS_2$ to 2L-$MoSe_2$/(iso-BA)$_2PbI_4$. (d) PL spectra of $ReS_2$/hBN/2L-$MoSe_2$/(iso-BA)$_2PbI_4$ (abbreviated as $ReS_2$/hBN/HS) and hBN/2L-$MoSe_2$/HS region under the excitation of 633 nm with a power of 200 μW at 78 K. (e) Power-dependent emission intensity of $ReS_2$/hBN/HS and hBN/2L-$MoSe_2$/HS and linear fittings. (f) Power dependence of enhancement factor in hBN/2L-$MoSe_2$/HS.

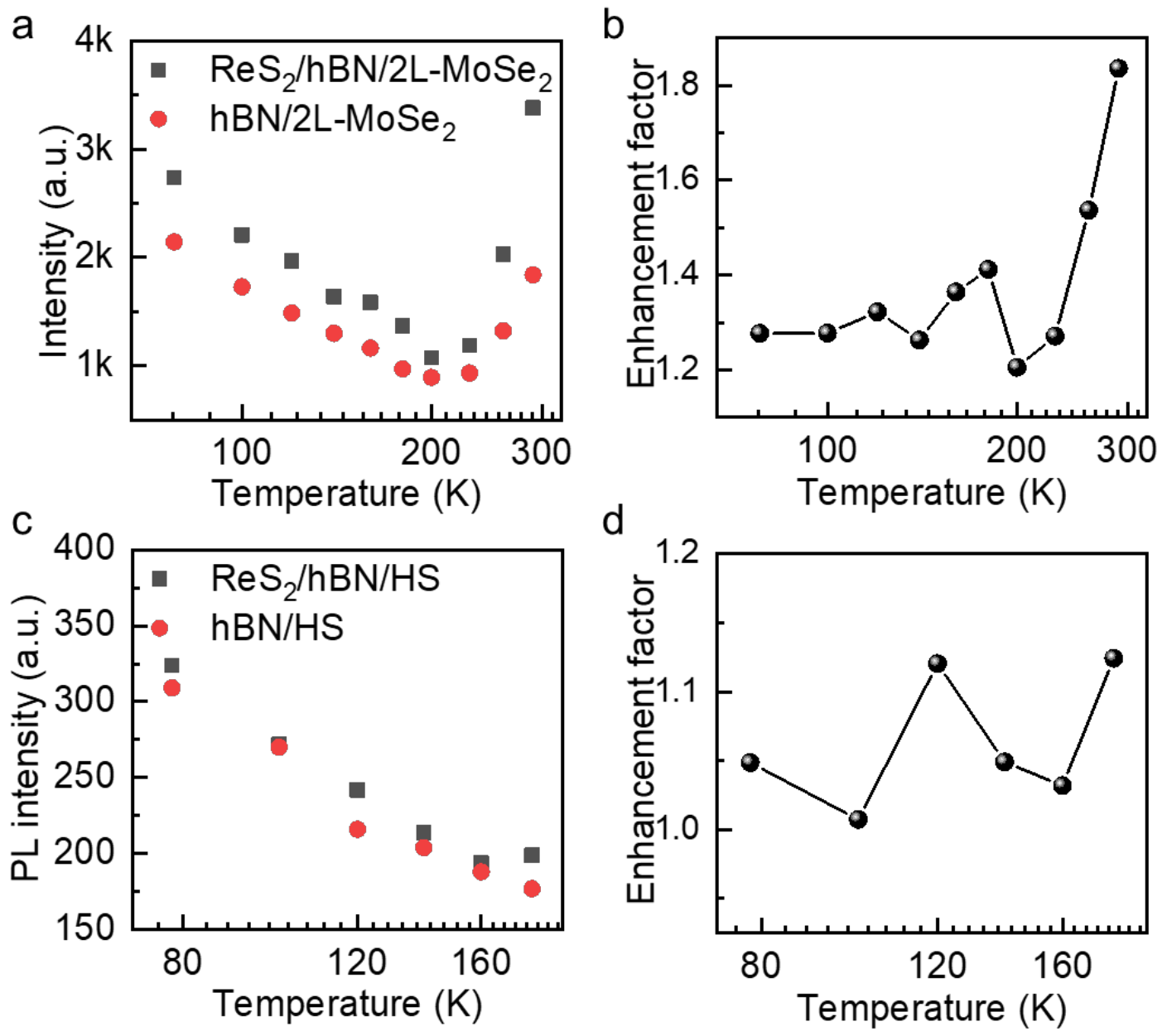

**Figure 3. Temperature dependence of PL enhancement.** Temperature-dependent PL intensity between different regions in $ReS_2$/hBN/2L-$MoSe_2$ (a) and $ReS_2$/hBN/HS (c) under the excitation of 633 nm with a power of 800 μW. Temperature-dependent evolution of PL enhancement factor of $ReS_2$/hBN/2L-$MoSe_2$ (b) and $ReS_2$/hBN/HS (d) under the excitation of 633 nm with a power of 2 mW.

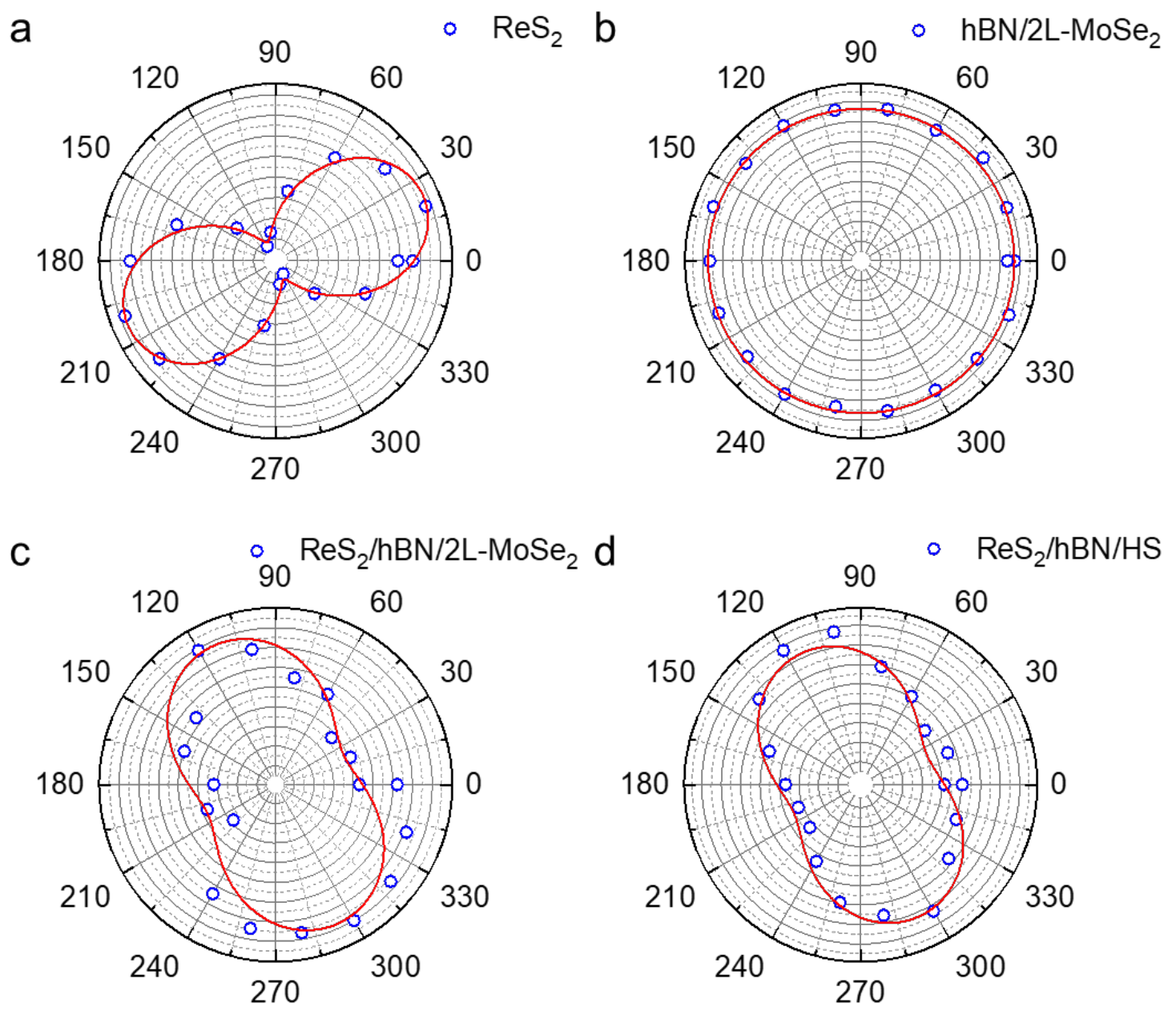

**Figure 4. Transfer of optical anisotropy accompanied by energy transfer.** Polar plots of polarization dependence of exciton emission intensity in $ReS_2$ (a), 2L-$MoSe_2$ (b), $ReS_2$/hBN/2L-$MoSe_2$ (c), and IXs in $ReS_2$/hBN/HS (d) under the excitation of 633 nm ith a power of 200 μW at 78 K.

# Supporting Information for

## Förster energy transfer boosts indirect anisotropic interlayer excitons in 2L-$MoSe_2$/perovskite heterostructures

Yingying Chen[1], Zihao Jiao[2], Haizhen Wang[2] and Dehui Li[2,3*]

*1 School of Science, Jimei University, Yinjiang Road 185, Xiamen 361021, China*

*2 School of Optical and Electronic Information, Huazhong University of Science and Technology, Luoyu Road 1037, Wuhan 430074, China*

*3 Wuhan National Laboratory for Optoelectronics, Huazhong University of Science and Technology, Luoyu Road 1037, Wuhan 430074, China*

*Correspondence to: Email: dehuili@hust.edu.cn.

This PDF file includes:

Supplementary Figs. 1 to 7

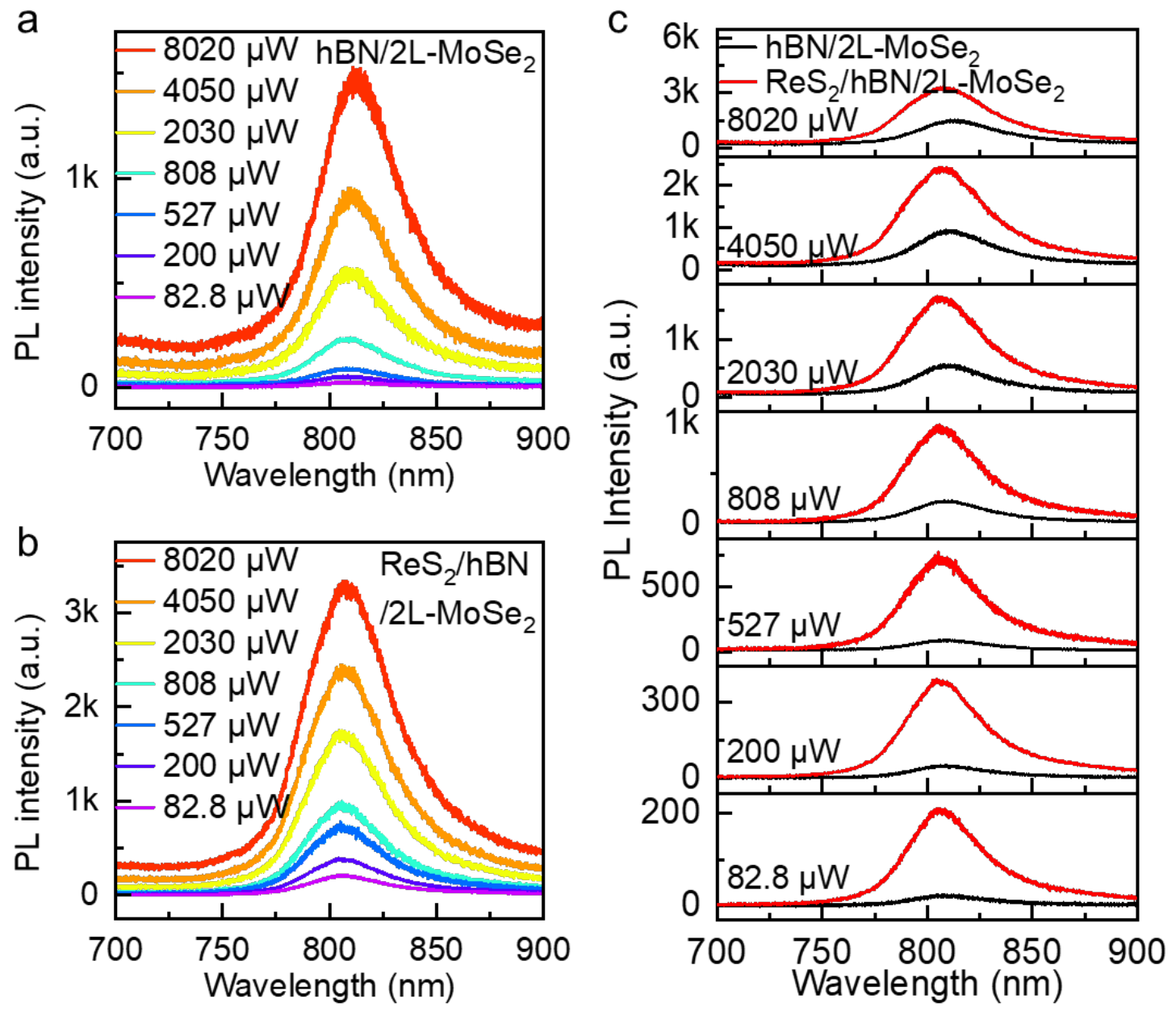

**Supplementary Fig. 1** Power-dependent PL spectra of hBN/2L-$MoSe_2$ (a) and $ReS_2$/hBN/2L-$MoSe_2$ (b) under the excitation of 633 nm at room temperature. (c) Detailed comparison of PL spectra between hBN/2L-$MoSe_2$ and $ReS_2$/hBN/2L-$MoSe_2$ under varying incident power.

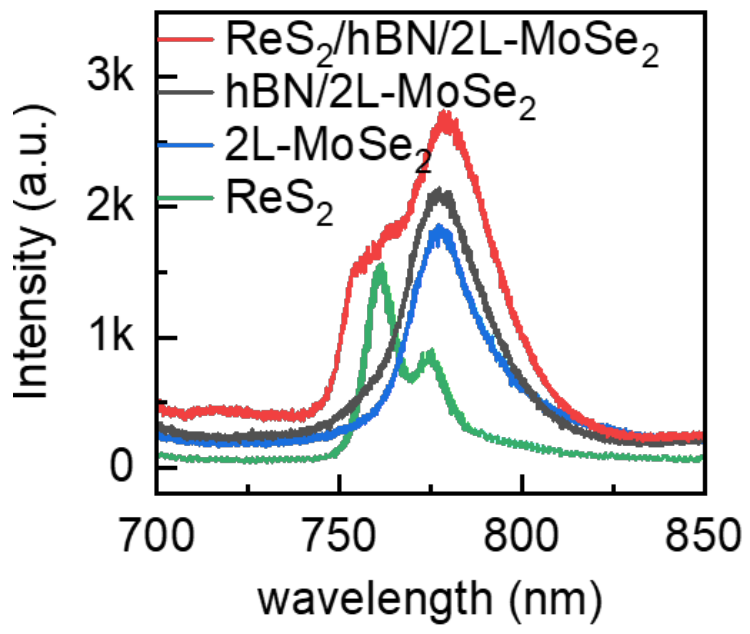

**Supplementary Fig. 2** PL spectra of the $ReS_2$/hBN/2L-$MoSe_2$ heterostructure and constituent individual regions under the excitation of 633 nm with the power of 800 μW at 78 K.

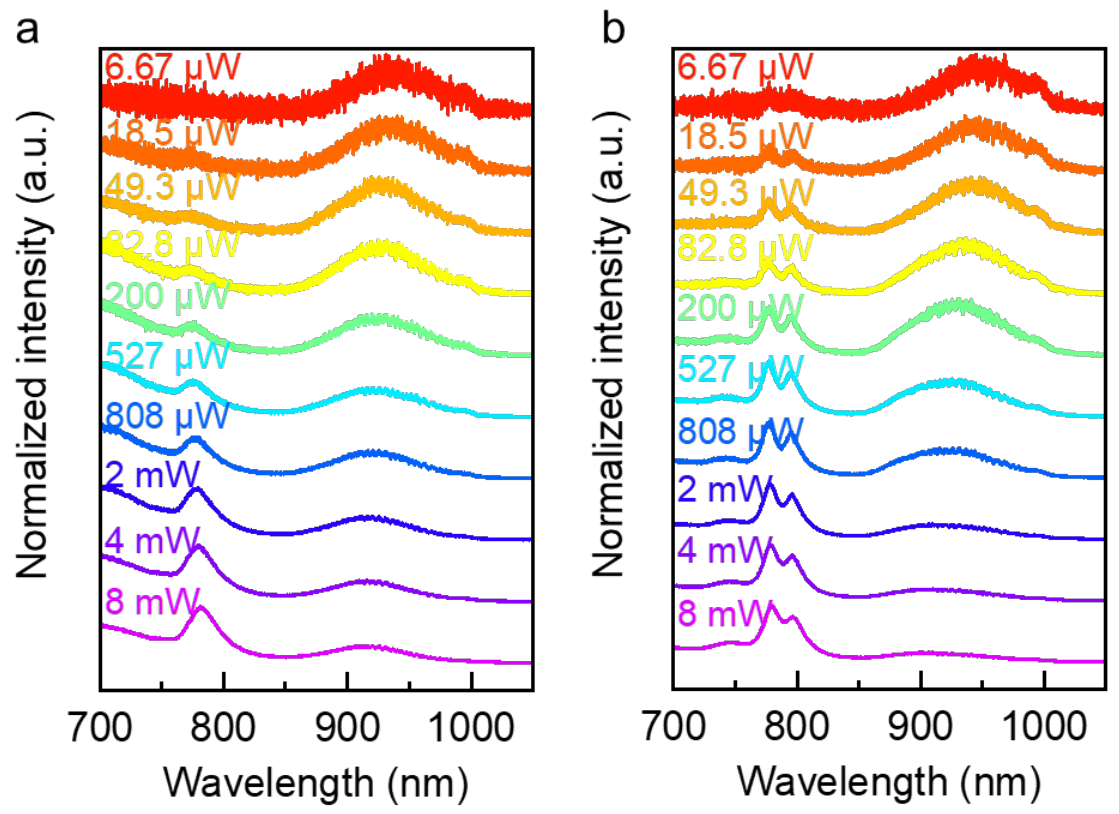

**Supplementary Fig. 3** Normalized power-dependent PL spectra of hBN/2L-$MoSe_2$/(iso-BA)$_2PbI_4$ (a) and $ReS_2$/hBN/2L-$MoSe_2$/(iso-BA)$_2PbI_4$ (b) under the excitation of 633 nm at 78 K.

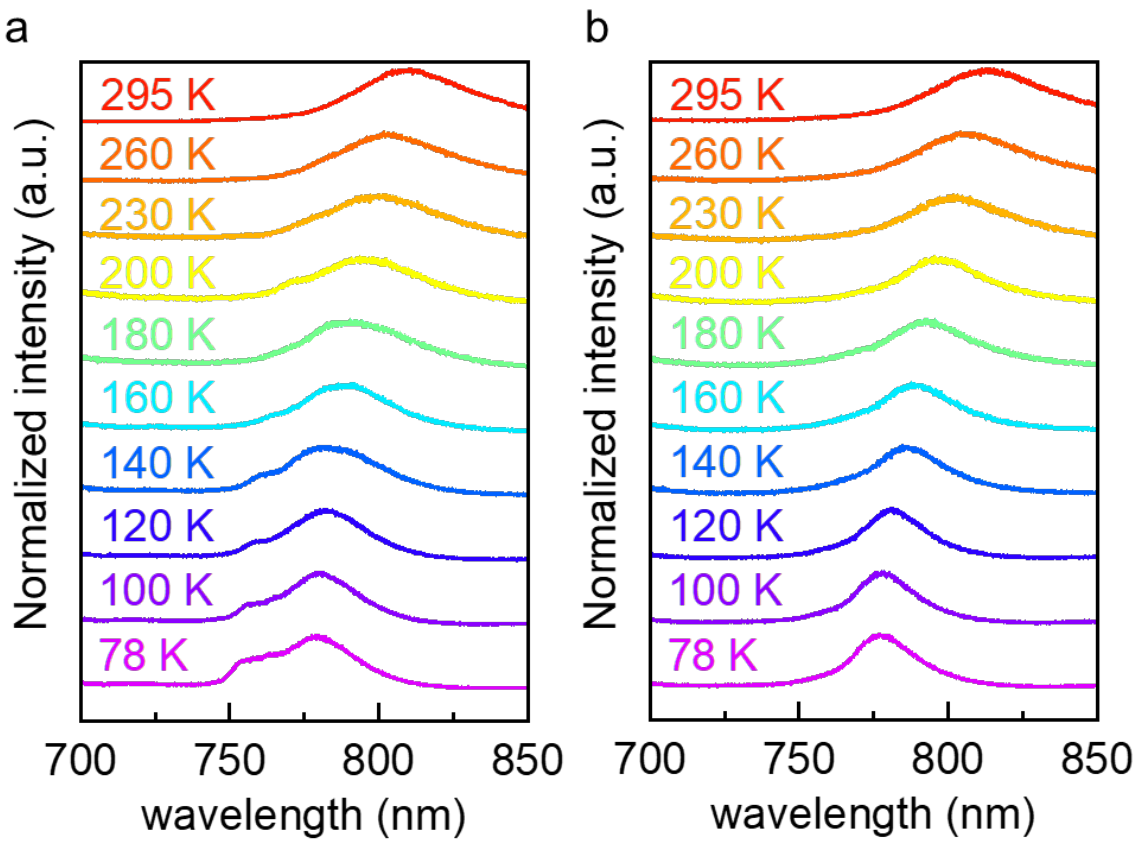

**Supplementary Fig. 4** Normalized temperature-dependent PL spectra of hBN/2L-$MoSe_2$ (a) and $ReS_2$/hBN/2L-$MoSe_2$ (b) under the excitation of 633 nm with a power of 800 μW.

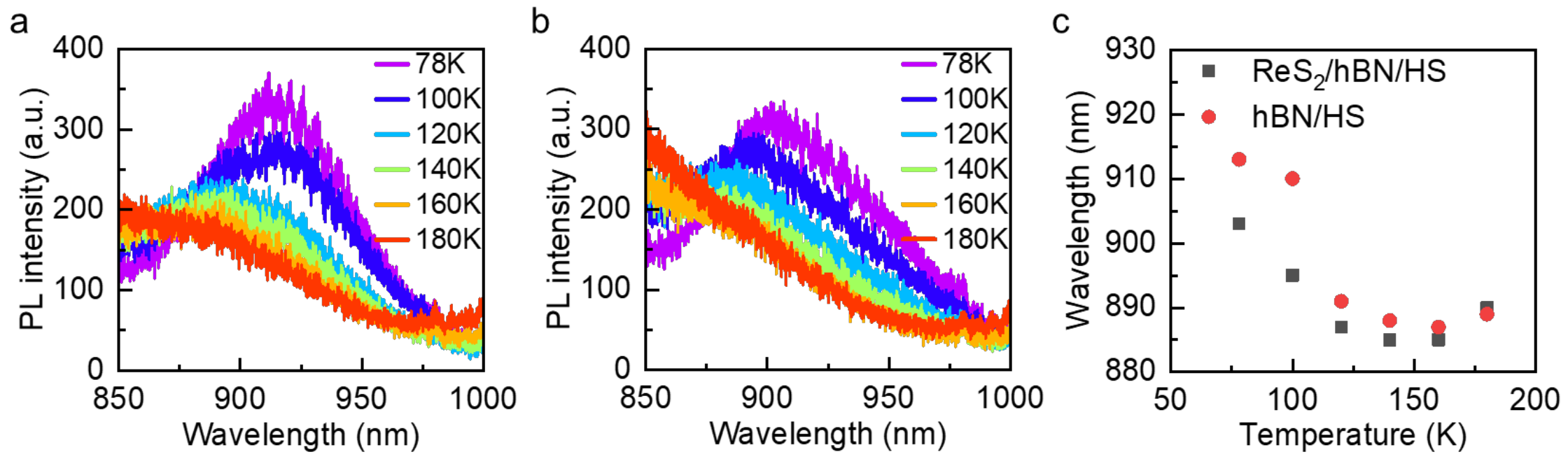

**Supplementary Fig. 5** Temperature-dependent PL spectra of hBN/2L-$MoSe_2$/(iso-BA)$_2$$PbI_4$ (a) and $ReS_2$/hBN/2L-$MoSe_2$/(iso-BA)$_2$$PbI_4$ (b) under the excitation of 633 nm with a power of 2 mW. (c) Central peak wavelength of hBN/2L-$MoSe_2$/(iso-BA)$_2$$PbI_4$ and $ReS_2$/hBN/2L-$MoSe_2$/(iso-BA)$_2$$PbI_4$ as a function of temperature.

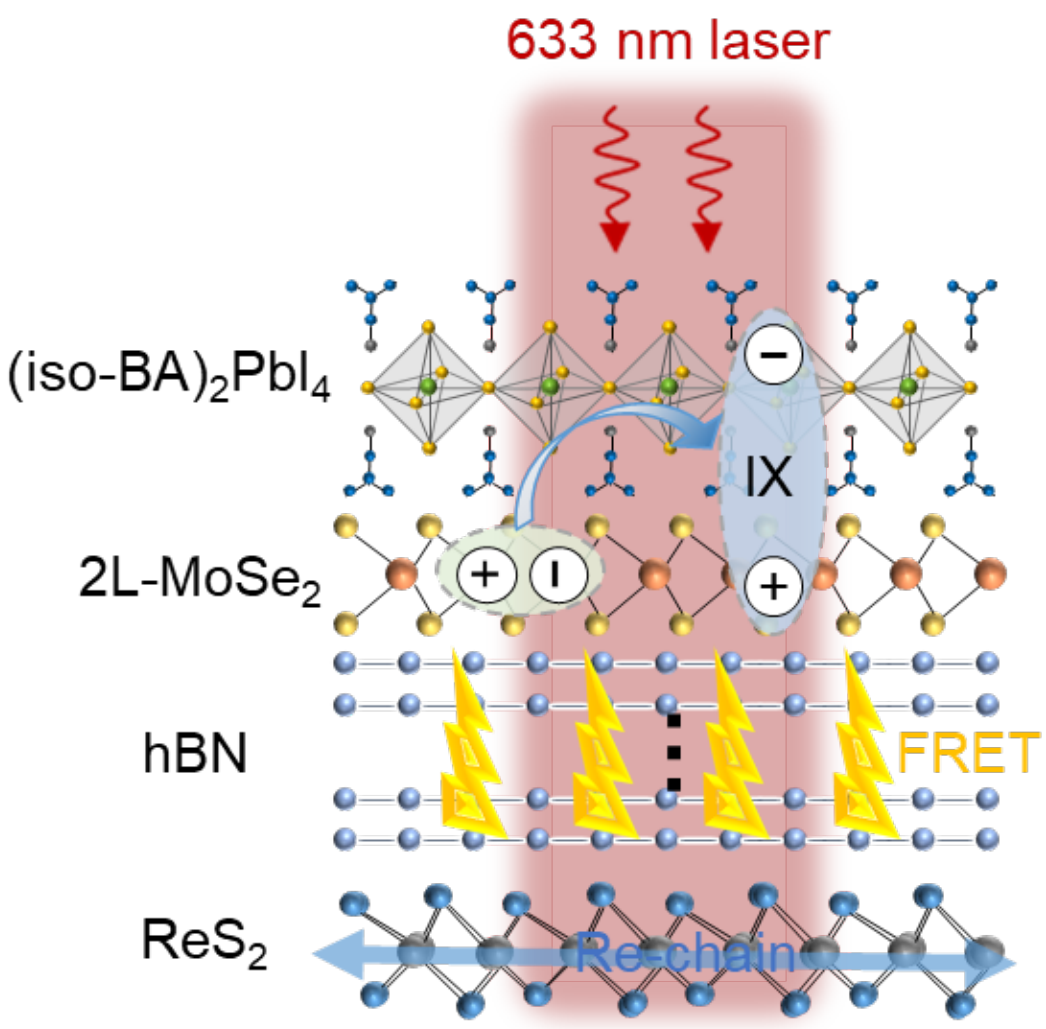

**Supplementary Fig. 6** Schematic illustration of the energy and charge transfer processes.

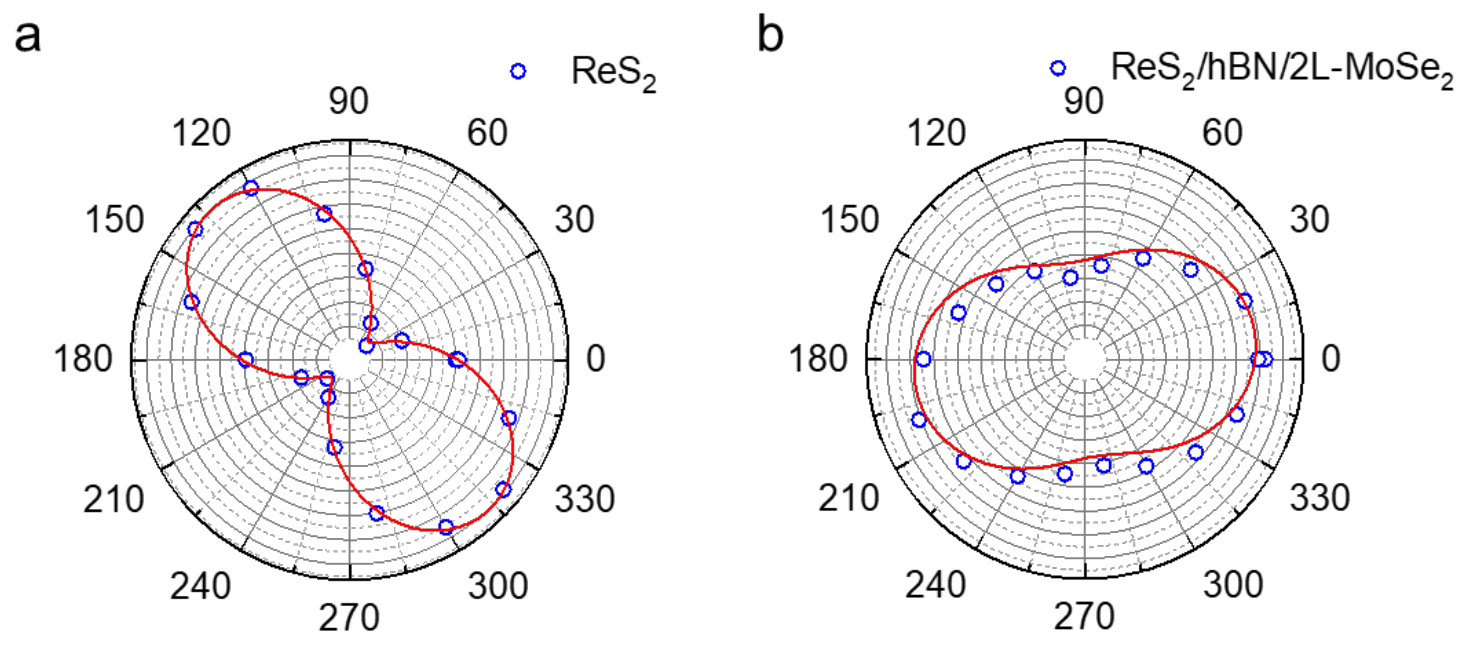

**Supplementary Fig. 7** Polar plots of linear polarization-dependent PL intensity of another $ReS_2$ (a) and $ReS_2$/hBN/2L-$MoSe_2$ (b) at room temperature, which shows a linear dichroism of ~ 2.0 and ~1.1, respectively.